\documentclass[preprint,12pt]{elsarticle}

\usepackage{amssymb}
\usepackage{amsmath}
\usepackage{graphicx}
\usepackage{times}
\usepackage{lineno}
\usepackage{upgreek}
\usepackage{lineno}
\usepackage{color,soul}

\usepackage{verbatim}

\journal{arXiv.org}

\begin{document}

\begin{frontmatter}



\title{The interplay of local chemistry and plasticity in controlling microstructure formation during laser powder bed fusion of metals}


\author[JHU]{Markus Sudmanns}
\ead{msudmanns@jhu.edu}
\author[NRL]{Andrew J. Birnbaum}
\author[IHPC,JHU]{Yejun Gu}
\author[NRL]{Athanasios P. Iliopoulos}
\author[NRL]{Patrick G. Callahan}
\author[NRL]{John G. Michopoulos}
\author[JHU]{Jaafar A. El-Awady\corref{cor1}}
\ead{jelawady@jhu.edu}

\cortext[cor1]{Corresponding author: }

\address[JHU]{Department of Mechanical Engineering, The Whiting School of Engineering, The Johns Hopkins University, Baltimore, MD 21218, USA}
\address[NRL]{United States Naval Research Laboratory, Washington DC, 20375, USA}
\address[IHPC]{Institute of High Performance Computing, A*STAR, Fusionopolis, 138632, Singapore}


\begin{abstract}
Additive manufacturing (AM) of metallic components promises many advantages over conventional powder or melting metallurgical manufacturing processes through high design flexibility across multiple length scales and precision coupled with an astonishing combination of mechanical properties.
Characterizing the relationship between microstructure and mechanical properties remains one of the major challenges for this novel technology.
A natural precursor is identifying the influence of the processing path on the developing microstructure.
We combine experimental studies of single track laser powder bed fusion (LPBF) scans of AISI 316L stainless steel, finite element analyses, and large-scale three-dimensional discrete dislocation dynamics simulations to provide a unique understanding of the underlying mechanisms leading to the formation of heterogeneous defect structures in additively manufactured metals.
Our results show that the interruption of dislocation slip at solidification cell walls is responsible for the formation of cellular dislocation structures, highlighting the significance of solute segregation for plastic deformation of additively manufactured components.
This work provides a mechanistic perspective on heterogeneous microstructure formation and opens the potential for a reliable prediction of the resulting mechanical properties of additively manufactured parts.

\end{abstract}


\end{frontmatter}

\section{Introduction}

To overcome the design and flexibility constraints imposed by conventional manufacturing, additive manufacturing (AM) of metallic components has evolved as a technique for building three-dimensional (3D) components, point-by-point, and layer-by-layer, directly from a digital model \cite{Herzog2016d,DebRoy2018f}.    
One such technique is laser powder-bed fusion (LPBF)  which uses high energy laser to melt and re-solidify metal powder to build complex components with high resolution \cite{Herzog2016d,Frazier2014b}.
Besides the inherent advantages over conventional powder or melting metallurgical manufacturing, remarkably different mechanical properties are usually observed for additively manufactured alloys as compared to conventionally manufactured ones, depending on the processing parameters \cite{lewandowski2016metal}, scan path \cite{Suryawanshi2017,CHEN201745}, heat-treatment \cite{YADOLLAHI2017218}, or even build orientation \cite{lewandowski2016metal, YADOLLAHI201714,YADOLLAHI2017218}.

In this context, additively manufactured austenitic AISI 316L stainless steel (SS) has received significant attention given its ubiquitous applicability \cite{Wang2018,Liu2018,Birnbaum2019,Bertsch2020}.
Significantly improved strength is observed as compared to conventionally manufactured components, sometimes without the traditional, concomitant sacrifice in ductility  \cite{Wang2018,Chen2019,Liu2018}.
Microstructural characterizations of as-build LPBF samples reveal a heterogeneous microstructure consisting of characteristic cellular dislocation substructures and chemical segregation in AISI 316L SS \cite{Wang2018}, multiple principle element alloys \cite{wu2018nanosized}, nickel-base superalloys \cite{zhang2017homogenization,vilaro2012microstructural}, and Al-Si alloys \cite{wu2016microstructure}.
These non-equilibrium microstructures are typically identified as being responsible for the unprecedented improvement in the mechanical properties \cite{Krakhmalev2018,Bertsch2020,Voisin2021}.
However, the relationship between chemical environment, microstructure, and mechanical properties is poorly understood, which prevents a reliable prediction and tailoring of mechanical properties of LPBF manufactured parts.
As AM is being developed towards a new paradigm of manufacturing \cite{APPLEYARD2015285}, predictive capabilities of the linkage between process-microstructure-properties for additively manufactured materials is becoming more of a necessity.
Designing the microstructure is one way to control the desired mechanical properties, requiring, however, identifying the influences of manufacturing on the microstructural evolution.

The influence of observed chemical segregation on the formation of dislocation substructures observed in LPBF-AISI 316L SS remains unclear and is still subject of debate \cite{Birnbaum2019,Bertsch2020,qiu2018comprehensive}.
The predominant theory attributes the existence of dislocation substructures to rapid cellular solidification driven by micro-segregation due to rejection of ferrite-stabilizing elements such as Molybdenum (Mo) and Chromium (Cr) into the melt during solidification, followed by the coincident formation of substructural interfaces \cite{Saeidi2015,Prashanth2017}.
However, other studies report cellular dislocation substructures in a homogeneous chemical environment \cite{Birnbaum2019, qiu2018comprehensive} even in as-built pure metals such as copper \cite{Wang2020f}.
Cooling rates, thermo-mechanical residual stresses and thermal cycling are usually emphasized and suggested as alternative explanations \cite{Birnbaum2019,qiu2018comprehensive,Wang2020f}.

A connection between dislocation microstructure and mechanical properties of LPBF manufactured parts can only be achieved by providing a comprehensible explanation of the observed microstructural evolution in light of the thermal and chemical environment during LPBF processing on a mechanistic level. 
Due to the characteristic spatial and temporal process regimes, using a purely experimental approach imposes limitations on the ability to track microstructure evolution on the level of individual dislocations in-situ, thus often necessitating a post-hoc approach for explanation. 
To properly represent the inherent complexities arising from the highly transient, multiscale, thermo-elastoplastic nature of the LPBF process, we employ an integrated experimental data-driven and computationally guided approach.
Three-dimensional (3D) Discrete Dislocation Dynamics (DDD) simulations are used to provide a unique understanding of the influence of solidification induced solute segregation on the evolution of dislocation ensembles.
By considering the interplay between local chemistry and dislocation plasticity during the cool-down phase after single track LPBF scans in AISI 316L SS, we capture the fundamental microstructural processes responsible for the formation of dislocation substructures, which are not accessible to experiments.

\section{Materials and Methods}

For a combined experimental and simulation approach, various challenges arise. First, determining the essential physical quantities that are sufficiently representative of the underlying LPBF process, and second, capturing dislocation mobility and interaction in a physically meaningful way in a challenging thermal and chemical environment that involves high temperatures, thermal induced residual stresses and solidification-induced segregation of alloying atoms. Therefore, experimental studies of single track LPBF scans in 316 SS are augmented by macro-scale finite element (FEM) analyses of the transient thermo-mechanical stresses induced and DDD simulations of the microstructure evolution during the cool-down phase that include the local, spatially varying chemical environment. The experimental and computational methods are discussed in the following subsections.

\subsection{Experimental methodology}
All single track experiments conducted here were generated on a GE/Concept Laser M2 LPBF system using a continuous wave laser with a power of $P = 370$\,W, laser scan speed of $V = 900\,$mm$/$s and laser spot size $d = 160\,\upmu$m. The concept Laser CL-20ES AISI 316L stainless steel powder was used with approximate composition Cr-17.5 at.\% Ni-11.5 at.\% Mo-2.3 at.\% Mn-1.0at.\% Si-0.5at.\% Fe Balance (P, C, S $<$ 0.03 at.\%). All single tracks were processed directly on a 5 mm thick AISI 316L base-plate, with a composition similar to that of the powder. 
The as purchased base-plate had an average grain size of $\approx 10\,\upmu$m. 
It was annealed at $1200^\circ$C for 24 hours to promote grain growth, and thus, reduce the influence of grain boundary interactions on the observations.
After annealing, the average grain size was $\approx 300\,\upmu$m. 
All cross sections were cut with a diamond saw, and subsequently grounded with successively finer grit from 320 to 1200 SiC paper. This was followed by polishing via 9\,$\upmu$m, 3\,$\upmu$m, and 40\,nm colloidal silica suspensions. Samples were then etched for 30 seconds in dilute oxalic acid (10\%). 
Scanning electron microscopy was performed on a JEOL JSM 7001F scanning electron microscope at 10 kV. Electron backscatter diffraction-based (EBSD) crystallographic indexing was performed on a JEOL JSM 7001F scanning electron microscope at 30 kV tilted 70$^\circ$ with respect to the incident electron beam. An EDAX EBSD camera was used with a 0.3 $\upmu$m step size and 4 ms acquisition time. An FEI Scios dual-beam focused ion beam with a Ga ion source operating at 30 kV was used to prepare thin foil samples for analysis. Scanning transmission electron microscopy (STEM) bright field (BF) imaging, and energy dispersive X-ray spectroscopy (EDS) mapping was performed using an aberration-corrected JEOL JEM-ARM200CF operating at 200 kV. 
Transmission Electron Microscopic (TEM) images were collected using a Tecnai G30 in BF STEM mode at 300 kV from a grain incorporating both segregated and unsegregated domains, see supplementary Fig.\,S1. 
The unsegregated TEM image was collected in a systematic row condition using a $(111)$ type Bragg condition.
The segregated TEM image was collected in a systematic row condition aligned to a $(022)$-type Bragg condition

\subsection{Finite element model and simulations setup}
\label{sec:FEM}

FEM simulations were conducted using the commercial FEM suite ABAQUS v.2020 to analyze the temperature profile and formation of residual stresses during LPBF processing using the experimental build conditions.
For the purpose of this study on a single track LPBF scan, a ballpark estimate of the residual stress magnitude and profile during cool-down is sufficient.
Therefore, various simplifying assumptions were employed.

First, the LPBF process in the FEM model was simplified by an isotropic thermo-mechanical model where the laser heat input was modeled by a direct uniform surface heat flux.
The intensity and size of the heat flux in the FEM simulation was adjusted such that the resulting melt pool size and shape was sufficiently equivalent to the experimentally observed re-solidified area limited by the fusion-zone boundary.
The heat source moves along the scanning direction with a velocity of 900 mm/s over a length of 3\,mm.

Second, we assumed a fully solidified homogeneous material as an initial condition, neglecting the influence of the powder bed.
Thus, the temporal evolution of the temperature and stresses induced by the process were taken from a region within the remelted baseplate.

Third,  a simple isotropic elastic-perfectly plastic material model was used, where all thermal and mechanical material parameters are temperature-dependent \cite{ChaudhuryAtlas, desu2016mechanical, byun25854mechanical}, as shown in supplementary Fig.\,S2.
A more accurate prediction of the stress profile at different temperatures could be achieved with a more advanced plasticity model in the FEM simulations. However, our preliminary analyses indicate that the current conclusions of the microstructure formation after single track LPBF do not change for different plasticity models. Nevertheless, such an advanced plasticity model could be especially valuable for investigating the effect of thermal cycling induced by multiple scan paths.

Finally, in the interest of simplicity we refrained from modeling the phase transformation associated with melting and re-solidification of the material and neglected effects of chemical segregation.
Instead, we relied on a simple melting and resolidification model built into ABAQUS, which removes the equivalent plastic strain when the temperature of a material point exceeds a defined temperature. Since we are only interested in an approximate estimate of the stress state and temperature profile induced by the LPBF process, this temperature was defined as 1700K (in the following referred to as melting temperature), which lies between the solidus and liquidus temperature (about 1680K and 1710K respectively) of AISI 316L stainless steel.
The mechanical boundary conditions are fixed boundary conditions on the bottom surface, $Z = 0$, (i.e., displacements in all directions are set to zero: $u_x=u_y=u_z=0$), see Fig.\,\ref{fig:Figure_2}a for the coordinate system.
Additionally, due to symmetry, only half of the system was modeled using the $XZ$-plane at $Y=0$ as the mirror plane, thus, the displacement on the side surface below the laser input $(Y=0)$ are set to zero in the $Y$-direction $(u_y = 0)$. 
All remaining surfaces are traction free.

\subsection{Discrete dislocation dynamics simulations setup}
\label{sec:DDD_method}

Large scale three-dimensional (3D) Discrete Dislocation Dynamics (DDD) simulations allow for a detailed study of the evolution of the dislocation microstructure in the cool-down phase of LPBF using processing conditions and accounting for the representative chemical microstructure. 
All DDD simulations conducted here employ an in-house version of the 3D DDD open-source code ParaDiS \cite{Arsenlis2007f}. 
The open source code was significantly modified in-house to avoid any artificial non-planar dislocation glide or collision events in FCC crystals as well as to incorporate atomistically-informed cross-slip mechanisms as described in \cite{Hussein2015}.

To reproduce the correct dislocation mobility in AISI 316L stainless steel, the stress induced by the alloying atoms on the dislocations as they glide in the FCC lattice have to be accounted for. 
This also includes the effect of spacial variations in the chemical composition.
To account for the solute effect on the dislocation critical resolved shear stress (CRSS) the theoretical model developed by Varvenne \textit{et al.} for a dislocation segment gliding in a single phase alloy with an arbitrary composition was utilized \cite{Varvenne2016g, Varvenne2017}. This approach was recently implemented into the 3D DDD framework to model dislocation evolution in multi-principle element alloys  \cite{SUDMANNS2021117307}.
The interaction between a dislocation segment and a solute site can be quantified by an energy barrier $\Delta E_b$, which the dislocation must overcome after reaching the related zero temperature strength $\tau_\mathrm{c}^0$ (i.e. dislocation Peierls stress) \cite{Varvenne2016g}. By considering thermal activation, the CRSS of the dislocation can be described as follows \cite{LEYSON20123873, Leyson2016}:
\begin{equation}
    \tau_\mathrm{c}(T,\dot{\varepsilon}) = \begin{cases}
    \tau_\mathrm{c}^0\text{exp}\left(-\frac{1}{0.52}\frac{kT}{\Delta E_b}\text{ln}\frac{\dot{\varepsilon}_0}{\dot{\varepsilon}}\right) \qquad 0.2 < \tau_\mathrm{c}/\tau_\mathrm{c}^0 < 0.5\\
    \tau_\mathrm{c}^0\left(\frac{1}{0.52}\frac{kT}{\Delta E_b}\text{ln}\frac{\dot{\varepsilon}_0}{\dot{\varepsilon}}\right)^{-\frac{1}{0.54}} \qquad \tau_\mathrm{c}/\tau_\mathrm{c}^0 < 0.03
    \end{cases}
    \label{eq:solute_stress}
\end{equation}
where $T$ is the temperature, $k$ is Boltzmann's constant, $\dot{\varepsilon}$ is the experimental strain rate, and $\dot{\varepsilon}_0 = 10^{4}\text{s}^{-1}$ is a reference strain rate \cite{Varvenne2016g}. Following Leyson and Curtin \cite{Leyson2016}, we separate a medium temperature regime between $0.2 < \tau_\mathrm{c}/\tau_\mathrm{c}^0 < 0.5$ from a high temperature regime below $\tau_\mathrm{c}/\tau_\mathrm{c}^0 < 0.03$, where an assumption has to be made for the transition between both regimes. To avoid a discontinuity, we approximate this transition by a Gaussian cumulative distribution function of a normal distribution. Then, the energy barrier and the dislocation Peierls stress can be expressed as \cite{Varvenne2016g}:
\begin{equation}
\tau_\mathrm{c}^0 = 0.051\alpha^{-\frac{1}{3}}\upmu(T)\left(\frac{1+\nu(T)}{1-\nu(T)}\right)^{\frac{4}{3}}f_1(w_c)\times\left[\frac{\sum_i c_i \Delta\bar{V}^2_i}{b^6}\right]^{\frac{2}{3}}
\label{eq:tauzero}
\end{equation}
and
\begin{equation}
\Delta E_b = 0.274\alpha^{\frac{1}{3}}\upmu(T)b^3\left(\frac{1+\nu(T)}{1-\nu(T)}\right)^{\frac{2}{3}}f_2(w_c)\times\left[\frac{\sum_i c_i \Delta\bar{V}^2_i}{b^6}\right]^{\frac{1}{3}}
\label{eq:deltaeb}
\end{equation}
where $\upmu(T)$ is the shear modulus, $\nu(T)$ the Poisson ratio, $b$ is the Burgers vector magnitude, $f_1(w_c) = 0.35$ and $f_2(w_c) = 5.70$ are core coefficients according to literature \cite{Varvenne2016g}, and $\alpha=0.5$ is the isotropic line-tension parameter close to previous assumptions for screw dislocations \cite{Chu2020}. The term $\sum_i c_i \Delta\bar{V}^2_i$ in Eqs. (\ref{eq:tauzero}) and (\ref{eq:deltaeb}) quantifies the misfit induced by solute $i$ compared to the average alloy, where $\Delta\bar{V}_i$ is the misfit volume of each solute which is calculated from spin polarized first principle calculations of Fe$_{0.687}$Ni$_{0.187}$Cr$_{0.125}$ \cite{antillonStructuralPointdefectCalculations2021}.
The temperature-dependencies of the shear modulus and the Poisson ratio in Eq.\,(\ref{eq:tauzero}) and (\ref{eq:deltaeb}) have been compiled from literature \cite{ChaudhuryAtlas, desu2016mechanical, byun25854mechanical} and summarized in supplementary Fig.\,S2. 

In the framework of 3D DDD, the solute concentration dependent dislocation segment CRSS as given by Eq. (\ref{eq:solute_stress}) is converted to a 'solute force' on each dislocation node resisting the dislocation motion, from which an effective nodal force $\mathbf{f}_\text{eff}$ is calculated as explained in \cite{SUDMANNS2021117307}. The effective nodal force $\mathbf{f}_\text{eff}$ is related to nodal velocities $v$ by a dislocation mobility law in the form of $v = M(\mathbf{f}_\text{eff})$, where $M$ is a mobility function whose arguments contain the effective nodal force as well as a temperature-dependent dislocation drag coefficient $B(T)$.
For the drag coefficient we apply an estimation for stainless steels which has recently been obtained from molecular dynamics (MD) simulations for Fe$_{0.7}$Ni$_{x}$Cr$_{0.3 - x}$ \cite{Chu2020}, where we use $x=0.15$.

\begin{figure}[h]
    \centering
    \includegraphics[width=\textwidth]{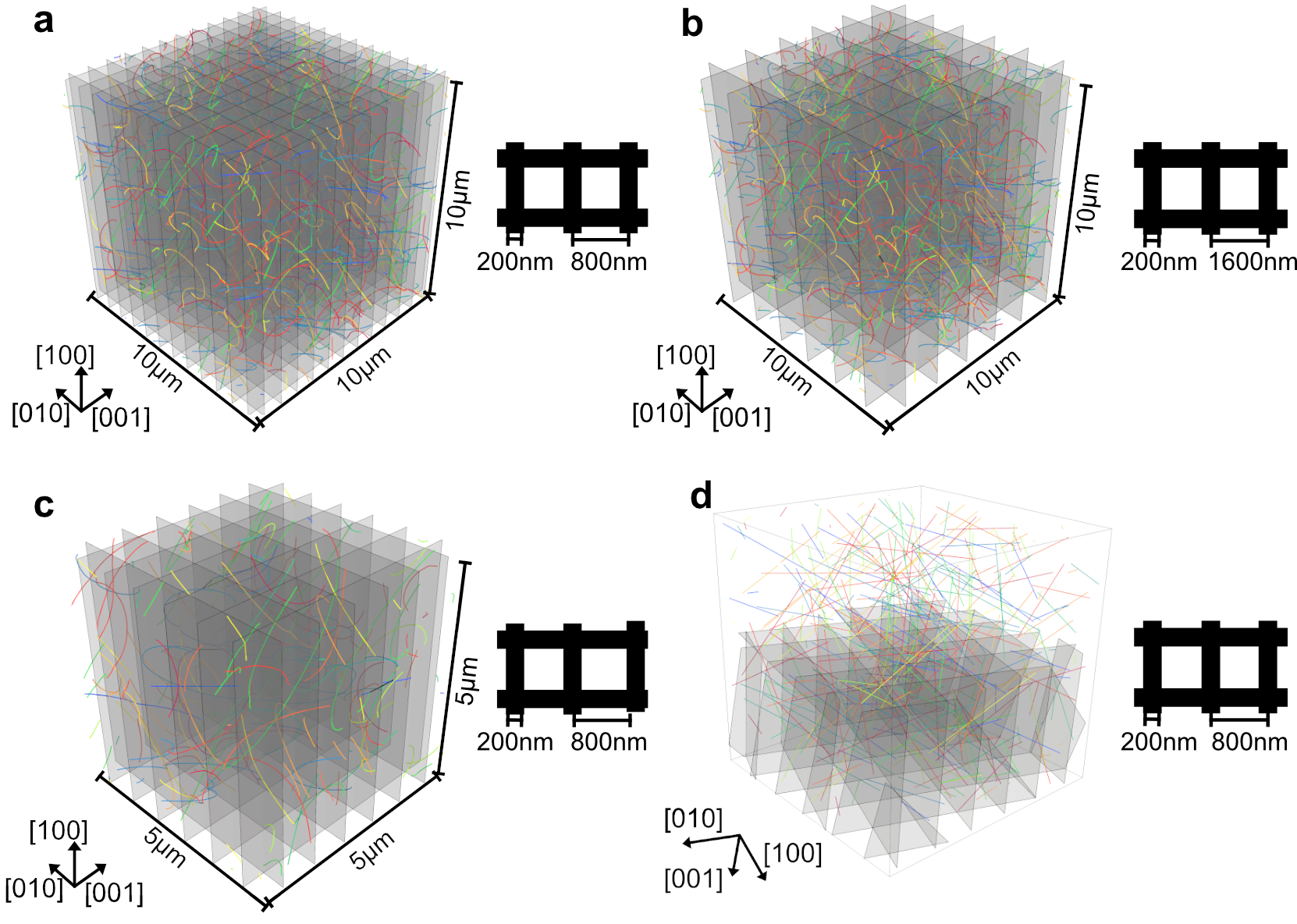}
    \caption{Schematic of the geometry of the solidification cell walls in the 3D DDD simulations with \textbf{a}, 800nm; and \textbf{b}, 1600nm cell wall spacing in a $10^3\,\upmu$m$^3$ simulation volume, \textbf{c}, the $5^3\,\upmu$m$^3$ DDD simulations with 800\,nm cell wall spacing, as well as, \textbf{d}, the geometry of the simulation with segregation walls in the lower half of the volume and the initial dislocation structure. Colored lines in \textbf{a}-\textbf{c} show the dislocation microstructure after the first simulation step, where different colors represent different slip systems. Grey planes indicate schematically the position of the center plane of the solute segregation walls.}
    \label{fig:Geometry}
\end{figure}

To mimic the temperature-dependent microstructure evolution in additively manufactured AISI 316L SS, we performed large-scale 3D DDD simulations at 1300K, 1100K, 900K and 700K in a $5^3\,\upmu$m$^3$ system volume containing cellular segregation according to experimental observations \cite{Wang2018, Liu2018, Depinoy2021}.
As a comparison, we also performed one simulation at 700K without cellular segregation.
Further, $10^3\,\upmu$m$^3$ 3D DDD simulations were conducted at 700K to allow for a more detailed study on the dislocation alignment in a larger representative volume.
All simulations containing cellular segregation incorporated a higher chemical concentration of Mo, Cr, and Mn in the cell walls as compared to the cell interior mimicking experimental observations. 
The details of the different composition are listed in Table \ref{tab:chemical_composition}.
A step-wise transition between cell wall and cell interior was assumed using a cell wall thickness of 200\,nm. 
The segregation cell walls are oriented parallel to the $[100]$ and $[010]$ plane with 800\,nm or 1600\,nm spacing according to the experimental observations \cite{Wang2018, Liu2018, Depinoy2021}, leading to cells elongated in the $[001]$ crystal direction, as shown by the schematics of the different DDD simulations in Fig.\,\ref{fig:Geometry}a-c.
The chemical segregation was assumed to form during solidification and is maintained during cool-down, following predominant explanations for the formation of the chemical cells by ejection of solutes with a high melting point at the solid-liquid interface \cite{Saeidi2015, Prashanth2017}.
For the simulations without cellular solidification, a uniform chemical composition for AISI 316L as shown in Table \ref{tab:chemical_composition} was assumed and homogeneously distributed throughout the volume.
The chemical composition used in all DDD simulations was chosen to approximately represent experimental measurements \cite{Birnbaum2019, Liu2018, Depinoy2021}.
\begin{table}[h]
\centering
\caption{Chemical composition in $($wt. $\%)$ of the solidification cell walls, cell interior and uniform chemical composition of AISI 316L stainless steel compiled from \cite{Liu2018, Birnbaum2019}}
\begin{tabular}{@{\extracolsep{4pt}}lccccc@{}}
\hline
Alloy & Fe & Ni & Cr & Mo & Mn\\
\hline
AISI 316L LPBF cell wall & 65.9 &  10.99 & 18.47 & 2.9 & 1.74\\
AISI 316L LPBF cell interior & 70.03 &  10.32 & 16.73 & 1.5 & 1.42\\
AISI 316L uniform & 67.7 &  11.5 & 17.5 & 2.3 & 1.0\\
\hline
\end{tabular}
\label{tab:chemical_composition}
\end{table}

The initial dislocation distribution in all DDD simulations consisted of Frank-Read sources with random orientation and length within a range of $1\,\upmu$m and $4\,\upmu$m, which corresponds to an initial dislocation density of $\approx 1.0 \times 10^{12}\,\mathrm{m}^{-2}$. 
The dislocations are placed on all possible 12 FCC slip-systems with a statistically equivalent initial dislocation density distribution, to avoid any bias in the evolution of dislocation density based on the initial dislocation distribution.
Although Frank-Read sources incorporate artificial pinning points, they are a reasonable assumption for the DDD simulations conducted here since the dislocation density increases by at least one order of magnitude.
This assumption is supported by recent 3D DDD simulations which show that in multi-slip orientations most dislocations are generated by dislocation multiplication mechanisms such as cross-slip or glissile junctions during evolution of dislocation networks and not due to a continued operation of static Frank-Read sources \cite{Stricker2018}.
For the sake of comparison, the same initial dislocation configurations were chosen for all $10^3\,\upmu$m$^3$ and $5^3\,\upmu$m$^3$ simulations, respectively, independent of the temperature range.
To mimic the microstructure evolution inside a single large grain away from the influence of the grain boundaries, we chose periodic boundary conditions for these DDD simulations.
The stress and temperature input for the DDD simulations was taken from the FEM simulation at a location within the melt pool with the temperature and all components of the stress tensor shown in supplementary Fig.\,S3.
The crystal orientation was chosen such that the $[100]$, $[010]$ and $[001]$ crystal axes align with the scanning (X), transverse (Y) and laser beam (Z) directions in the FEM simulation.

To provide a direct comparison with experimental results, we further performed DDD simulations which mimic the crystal orientation and solidification morphology as observed experimentally.
The crystal under investigation is rotated by the Euler-angles $\phi = -102.48^\circ$, $\theta = 162.62^\circ$ and $\psi = -171.60^\circ$ using the Z-X-Z convention with respect to the specimen coordinate system shown in Fig.\,\ref{fig:Figure_1}a.
The stress tensor obtained from the FEM simulation was applied upon its correct rotation. 
In this simulation, the single grain under investigation in DDD is represented by a $10^3\,\upmu$m$^3$ sized simulation volume which is separated into a segregated and unsegregated domain.
Here, we used free surfaces in order to avoid a periodic replication of the boundary between segregated and unsegregated domains. 
To minimize the influence of free surfaces on the results, we restricted the investigation on a sub-volume in the center of the simulation domain with $5^3\,\upmu$m$^3$.
The initial dislocation distribution consisted of Frank-Read sources with random orientation and length within a range of $1\,\upmu$m and $4\,\upmu$m, which corresponds to an initial dislocation density of $\approx 1.0 \times 10^{12}\,\mathrm{m}^{-2}$.

All DDD simulations shown in this study contain a Burgers vector of $b = 0.249$\,nm and a temperature-dependent shear modulus $\mu(T) = E(T)/(2(1+\nu(T)))$, using the elastic modulus $E(T)$ and Poisson's ratio $\nu(T)$ as shown in supplementary Fig.\,S2.

Two-dimensional (2D) spatial correlation functions were used to provide an identification and a quantitative analysis of periodic patterns in the dislocation microstructure. We calculated the autocorrelation function in the Fourier space by making use of the Wiener-Khinchin theorem, which relates the autocorrelation function and the Fourier transform \cite{wiener1930generalized, khintchine1934korrelationstheorie}. Further, we used the $5^3\,\upmu$m$^3$ 3D DDD simulation at 700K without pre-existing solute segregation to quantify the solute diffusion during cooldown and thus estimate the effect of segregation induced by solute diffusion. A detailed discussion of both techniques is given in the supplementary section S1.1 and S1.2.

\section{Results and Discussion}

\subsection{Experimental characterization of microstructures in single track LPBF scans}

\begin{figure}[htbp]
    \centering
    \includegraphics[width=\textwidth]{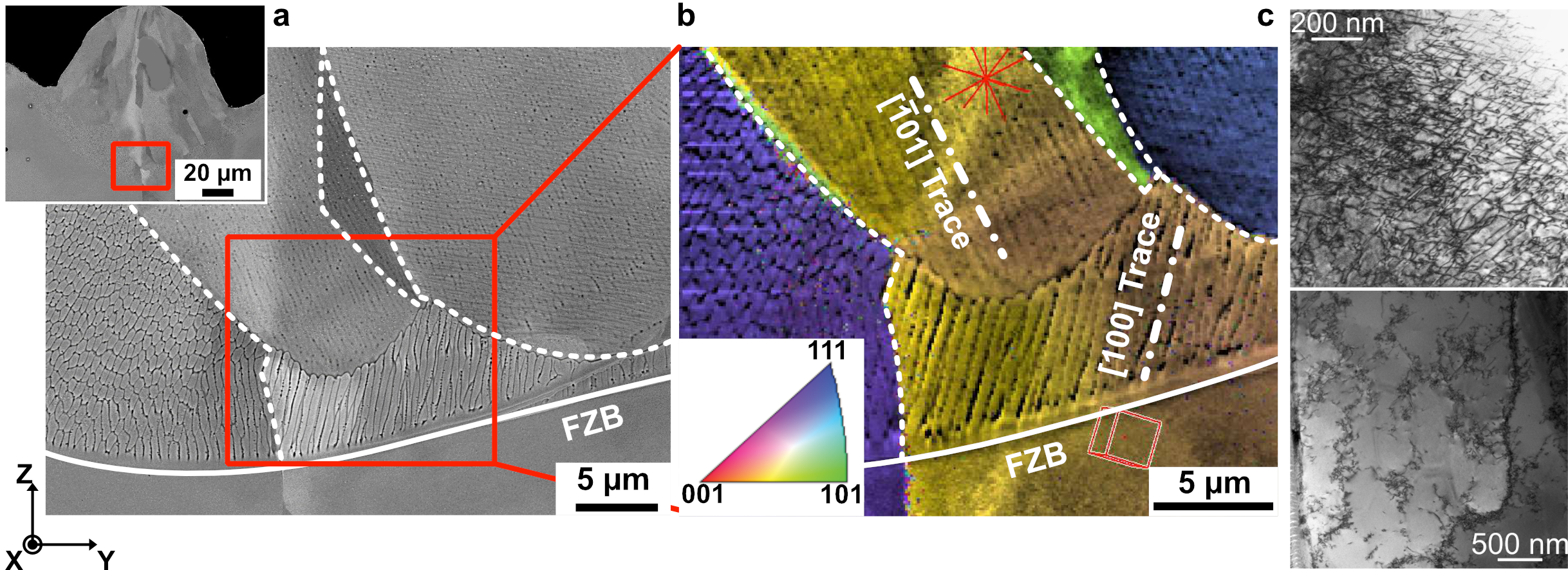}
    \caption{\textbf{Formation of different substructures within the same grain after a single-scan LPBF in AISI 316L stainless steel.}
    \textbf{a}, Magnified scanning electron microscopy (SEM) image of a cross section (inset) close to the fusion-zone boundary (FZB) indicated by the solid white line, which delimits the melt pool. The X-axis corresponds to the laser scanning direction, while the negative Z-axis corresponds to the laser beam direction. Dashed white lines indicate grain boundaries.
    \textbf{b}, Electron backscatter diffraction (EBSD) inverse pole figure (IPF) map corresponding to the SEM image in \textbf{a} showing crystallographic orientations of different visible structural traces in a single grain.
    \textbf{c}, Transmission electron microscopic (TEM) images from areas with $\left<110\right>$-trace (top) and $\left<001\right>$-trace (bottom).
   }
    \label{fig:Figure_1}
\end{figure}

Single track laser scans are the fundamental sub-unit of the LPBF process. 
Consequently, in order to allow for a targeted study of the relationship between processing conditions and microstructure formation without the influence of recurring heat input, single tracks were generated by a GE/Concept Laser M2 LPBF system using a single powder layer on a polycrystalline AISI 316L baseplate. This baseplate which was annealed prior to the laser processing to reduce the influence of grain boundary interactions.
Cross-sections perpendicular to the laser scanning direction have been extracted and analyzed by Scanning Electron Microscopy (SEM), Electron Backscatter Diffraction (EBSD) and Transmission Electron Microscopy (TEM).
Figure\,\ref{fig:Figure_1}a shows an SEM image of the cross-section of an as-built sample after etching with different surface patterns within different grains clearly observed. 
While the characteristics of these patterns clearly depend on the grain orientation, different patterns are also observed within a single grain as is evident from the elongated grain in the center of the magnified SEM image in Fig.\,\ref{fig:Figure_1}a.
This is also evident from the corresponding Inverse Pole Figure (IPF) map in Fig.\,\ref{fig:Figure_1}b.
In that particular grain, a structural alignment is observed parallel to the $[\bar{1}01]$ crystal direction (indicated by the upper white dash-dot line), which is distinctively different from adjacent features that exhibit a clear and distinct $[100]$-alignment (lower white dash-dot line). 
The latter structure can be characterized as solidification cells \cite{Liu2018, Bertsch2020, Voisin2021}. 
The TEM images below each pattern extracted from a section inside the grain (see location in Fig.\,S1) reveal distinctly different dislocation cell structures, as shown in Fig\,\ref{fig:Figure_1}c.
While irregular dislocation structures are observed in the $\left<110\right>$-structured domain (top), clear dislocation cell walls are evident in the region that exhibits solidification cells (bottom). 
Additionally, our earlier TEM-EDS analysis revealed that cellular segregation of Mo and Cr is only observed in regions that exhibited clear dislocation cell walls \cite{Birnbaum2019}. 
Consequently, two characteristic substructures can be identified: (i) A \textit{``solute segregation-induced"} domain consisting of dislocation and solute segregation subcells, and (ii) a \textit{``dislocation interactions-induced"} domain consisting of diffuse dislocation cellular structures that are chemically homogeneous.
From the relative locations of the two regions with respect to the fusion-zone boundary (FZB), it can be inferred that a sudden change in growth mode occurred upon resolidification whereby the growth was initially cellular, but may have become rapidly planar.
We note that the appearance of both features within the same grain to date has not been reported in LPBF-processed bulk samples, and thus, might be due to the promoted grain growth which was performed prior to LPBF processing, absence of cyclic heating in the single-track  scan, or other processing conditions. Nevertheless, single-track scans provide an ideal setting to study the microstructure formation during LPBF processing on a fundamental level using a combined experimental and computational approach. Furthermore, the microstructure features observed during single-track behavior is a precursor to microstructures observed after full bulk AM builds and gaining a fundamental understanding of the underlying deformation mechanisms that result in these microstructures in single-track scans is thus a necessity before being able to accurately predict the bulk AM microstructure.

\subsection{Evolution of the transient temperature and macroscopic residual stresses}

\begin{figure}[htbp]
    \centering
    \includegraphics[width=\textwidth]{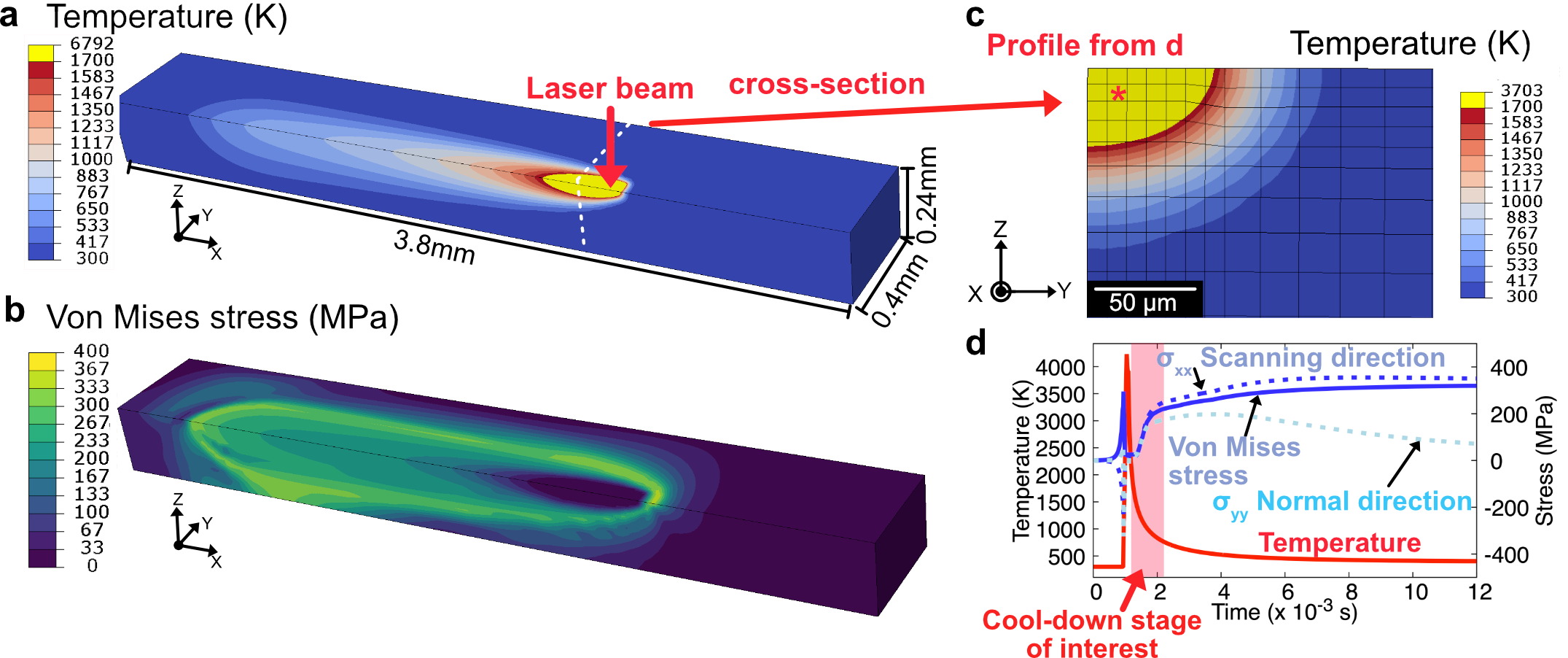}
    \caption{\textbf{Temperature profile and residual stresses during single track LPBF processing as predicted from FEM simulations.} 
    \textbf{a}, the predicted 3D contours of the temperature; and \textbf{b}, the von Mises stress profiles during single track LPBF processing. 
    Due to symmetry in the $Y$-direction, only half of the system is shown. Temperatures exceeding the melting temperature of 1700\,K (see section \ref{sec:FEM}) are shown in yellow.
    \textbf{c}, Contours of the predicted temperature on the cross-section shown in \textbf{a} by dashed lines. 
    \textbf{d}, The temperature, von Mises stress, and the stress components in the scanning and normal directions as a function of time at the point shown in \textbf{c} by an '*'.
    The X-axis corresponds to the laser scanning direction, while the negative Z-axis corresponds to the laser beam direction.}
    \label{fig:Figure_2}
\end{figure}

Prior to understanding the evolution of dislocation microstructure during the single track LPBF scans of AISI 316L, the formation of thermally induced stresses originating from the cool-down process must be evaluated. This is achieved through a set of thermally driven FEM simulations of the one way coupled thermo-mechanical problem under thermo-elasto-plastic constitutive conditions as explained in section \ref{sec:FEM}. 
The simulations were conducted using the commercial FEM suite ABAQUS incorporating temperature-dependent material parameters as summarized in supplementary Fig.\,S2. 
The temperature profile at an intermediate step during the LPBF processing is shown in Fig.\,\ref{fig:Figure_2}a, where, due to symmetry, only half of the system is shown. 
A shallow and elongated melt pool can be observed (shown in yellow) with temperatures decreasing rapidly after dropping below the melting temperature of 1700\,K (Fig.\,\ref{fig:Figure_2}d). The rapid temperature drop is accompanied by a significant rise in stresses inside the melt pool region as well as beyond the FZB. This is shown by the von Mises stress profile in Fig.\,\ref{fig:Figure_2}b for the time step corresponding to Fig.\,\ref{fig:Figure_2}a. 
The magnitude of the residual stresses exceeds 300\,MPa inside the re-solidified area upon cool-down, as shown by the time evolution of the temperature, von Mises stress, and the relevant components of the stress tensor in Fig.\,\ref{fig:Figure_2}d.
Furthermore, it is evident that the normal stresses parallel, $\sigma_{xx}$, and perpendicular, $\sigma_{yy}$, to the laser scanning direction are the dominant stress components that remain after re-solidification (see supplementary Fig.\,S3 for all stress tensor components).

These residual stresses are generated due to the constraint on contraction in the melt pool upon cooling, enforced by the unmelted material at the FZB and agree well in magnitude with experimental measurements of residual stresses in bulk LPBF AISI 316L SS samples \cite{simson2017residual}.
The predicted melt-pool geometry is in good qualitative agreement with the single track experimental observations, shown in Fig.\,\ref{fig:Figure_1}a.
Given the predicted rapid saturation in stress, the cool-down stage of interest for the subsequent analyses can be restricted to a small time interval following solidification that exhibits the steepest thermal stress gradients, as highlighted in Fig.\,\ref{fig:Figure_2}d.

\subsection{Temperature-dependency of dislocation structure formation during cool-down}

The highly transient LPBF process suggest a significant influence of temperature and residual stresses on the formation of the experimentally observed cellular dislocation structure.
In order to investigate this dependency, isothermal 3D DDD simulations have been conducted at 1300K, 1100K, 900K and 700K. 
We account for a physically meaningful representation of the dislocation evolution in AISI 316L SS via a temperature-dependent dislocation mobility law adapted from molecular dynamics (MD) simulations \cite{Chu2020} and a recently developed model for the local composition-dependent critical resolved shear stress (CRSS) \cite{SUDMANNS2021117307}.
The experimentally observed $\left<100\right>$-aligned and periodic spatial variation in chemical composition (local Mo and Cr enrichment) adapted from \cite{Liu2018,Birnbaum2019,Depinoy2021} are explicitly incorporated in the simulations, see Fig.\,\ref{fig:Geometry}c for the solidification cell wall geometry.
The thermo-mechanical residual stress evolution obtained from the FEM simulations (Fig.\,\ref{fig:Figure_2}d) serve as the foundation for the prediction of the dislocation structure evolution in these DDD simulations, which represents the microstructure as a snapshot during the cool-down stage. 
To allow for a meaningful comparison of the temperature-effect, we use the same initial dislocation configuration for each simulation and perform the investigation to similar levels of von Mises equivalent plastic strain $\varepsilon_\mathrm{pl}^\mathrm{VM}$.

\begin{figure}[htbp]
    \centering
    \includegraphics[width=\textwidth]{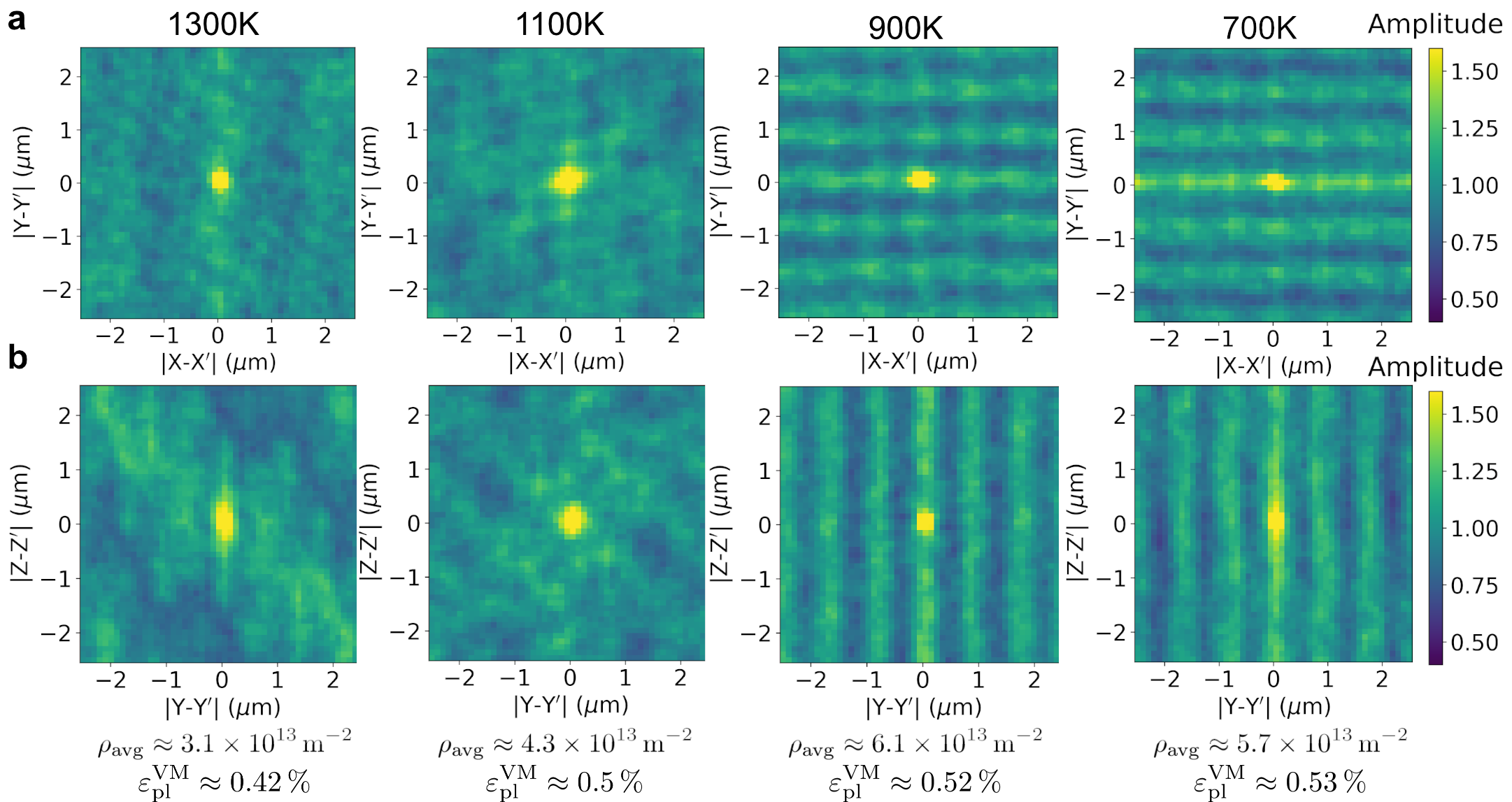}
    \caption{\textbf{Temperature-dependent solute segregation-induced dislocation structure formation after solidification and during cool-down.} 
    Autocorrelation functions of the dislocation structures predicted from $5^3\,\upmu$m$^3$ 3D DDD simulations at different temperatures (see supplementary Fig.\,S4). The $X$-, $Y$- and $Z$-axes correspond to the $[100]$, $[010]$ and $[001]$ crystal directions. 
    \textbf{a}, 2D contour-plots showing the amplitude of autocorrelation functions using the average local dislocation density in the $[001]$-direction (see supplementary Fig.\,S5), where only local densities that exceed the full volume average density by more than 15\% are considered.
    \textbf{b}, Similar to \textbf{a}, but for an average along the $[100]$-direction.
    }
    \label{fig:Figure_3}
\end{figure}

The predicted dislocation microstructure as an average local dislocation density map in the top (i.e. $[001]$ direction) and side  (i.e. $[100]$ direction) view of the 3D simulation cell is shown in supplementary Fig.\,S4.
An evaluation of the periodicity in the predicted dislocation microstructures at each temperature is shown by 2D contour-plots of the amplitudes of autocorrelation functions (see supplementary section S1.1) of the dislocation density averaged along the $[001]$ and $[100]$ direction, i.e. parallel and perpendicular to the segregation cells, as shown in Fig.\,\ref{fig:Figure_3}.
The amplitude of the autocorrelation function at 1300K and 1100K (left two columns) shows no obvious periodicity along any $\left<100\right>$ crystal direction. 
At lower temperatures, however, a periodic pattern in the $\left<100\right>$ crystal directions becomes more apparent, replicating the solidification cell spacing of 800\,nm in both the perpendicular and longitudinal directions.

These DDD simulations show a temperature dependence on the formation of the periodic dislocation structures with clear cellular dislocation substructures beginning to form at temperatures below 1100K. 
At higher temperatures, the dislocation structure (supplementary Fig.\,S4) resembles an irregular heterogeneous dislocation microstructure, which is commonly observed in the absence of solidification cells (e.g. Fig.\,\ref{fig:Figure_1}c, top). 
This indicates a dominance of dislocation multiplication mechanisms such as cross-slip over dislocation-solute interaction at high temperatures, whereas the solute induced resistance to dislocation glide becomes relevant with progressing cool-down.

\subsection{The effect of cellular segregation on the formation of dislocation structures}

The influence of cellular solidification (directional solidification driven by micro-segregation) on the formation of the dislocation structure at 700K is further analyzed by isothermal 3D DDD simulations as explained in section \ref{sec:DDD_method}.
We consider $\left<100\right>$-aligned solidification cells with either 800\,nm or 1600\,nm cell wall spacing to mimic experimental observations \cite{Wang2018, Liu2018, Depinoy2021} (see Fig.\,\ref{fig:Geometry}a,b for the geometry) and compare with a 3D DDD simulation in the absence of cellular segregation (i.e. chemically homogeneous).

\begin{figure}[htbp]
    \centering
    \includegraphics[width=\textwidth]{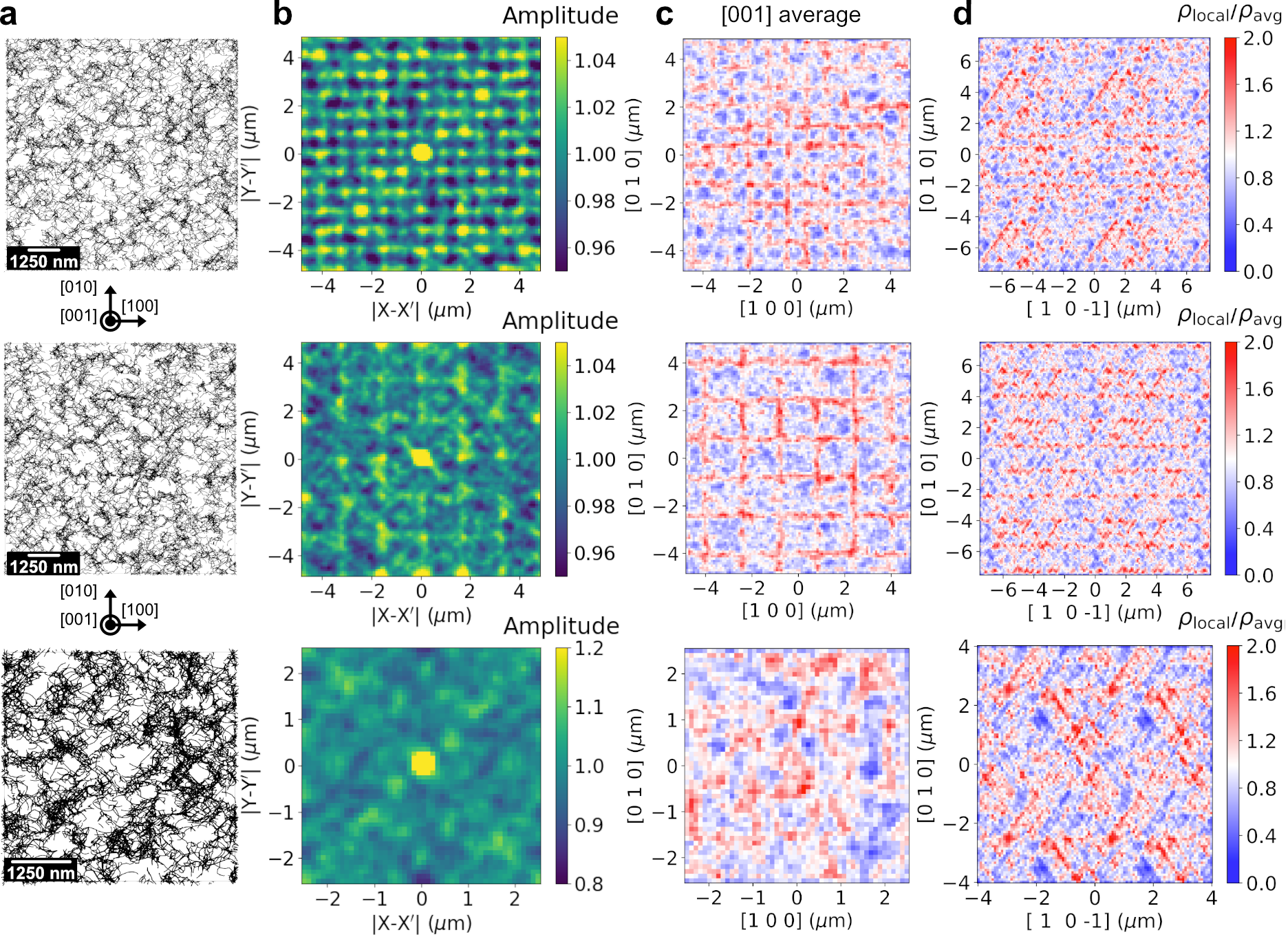}
    \caption{\textbf{Effect of cellular solidification on dislocation structure alignment.} 
    Dislocation structures predicted by $10^3\,\upmu$m$^3$ 3D DDD simulations at 700K with 800\,nm at $\rho_\mathrm{avg}\approx9.6\times10^{13}\,\mathrm{m}^{-2}$ and $\varepsilon_\mathrm{pl}^\mathrm{VM}\approx1.1\%$ (top row), and 1600\,nm at $\rho_\mathrm{avg}\approx1.0\times10^{14}\,\mathrm{m}^{-2}$ and $\varepsilon_\mathrm{pl}^\mathrm{VM}\approx1.07\%$ (center row) solidification cell sizes, and a $5^3\,\upmu$m$^3$ 3D DDD simulation in the absence of cellular segregation at $\rho_\mathrm{avg}\approx1.46\times10^{14}\,\mathrm{m}^{-2}$ and $\varepsilon_\mathrm{pl}^\mathrm{VM}\approx1.7\%$ (bottom row). 
    \textbf{a}, The predicted dislocation structures in 400\,nm thick sections extracted from the center of each simulation cell normal to the $[001]$ direction (see coordinate system for reference). 
    \textbf{b}, 2D contour-plots showing the amplitude of the autocorrelation functions using an average local dislocation density map extracted from the sections shown in \textbf{a}, where the $X$- and $Y$-axes correspond to the $[100]$ and $[010]$ crystal directions, respectively.     
    \textbf{c}, Average local dislocation density $\rho_\mathrm{local}$ relative to the full volume average density $\rho_\mathrm{avg}$ along the $[001]$ direction. 
    \textbf{d}, Similar to \textbf{c}, but averaged along the $[101]$ direction.
    }
    \label{fig:Figure_4}
\end{figure}

In the DDD simulations with 800 nm and 1600 nm cell sizes, a dense dislocation microstructure is formed driven by the large residual stresses, as shown in Supplementary Movies M1 and M2.
``Digital'' TEM-like lamellae with 400\,nm thickness have been extracted from the center of each 3D DDD simulation normal to the $[001]$ crystal direction (i.e. corresponding to the $Z$-direction in Fig.\,\ref{fig:Figure_2}) and are shown in Fig.\,\ref{fig:Figure_4}a. 
With cellular solidification (top and center rows), a fairly periodic cellular dislocation structure is observed containing dislocation alignment parallel to the $\left<100\right>$ directions. 
Additionally, the presence of a second, co-existing structural alignment parallel to $\left<110\right>$ directions is also observed (see coordinate system for reference). 
In contrast, a section extracted from the DDD simulation without cellular segregation (bottom row) reveals a microstructure with diffuse dislocation cell structures aligned mostly parallel to the $\left<110\right>$ crystal orientations, which is consistent with the features observed via etching in oxalic acid, Fig. \ref{fig:Figure_1}. 

To identify structural dislocation patterns, the amplitude of 2D autocorrelation functions for all digital TEM-foils in Fig.\,\ref{fig:Figure_4}a are shown as contour-plots in Fig.\,\ref{fig:Figure_4}b. 
For both simulations containing cellular segregation, a periodic pattern in both the $X$ and $Y$ directions are evident with a periodic spacing corresponding to the chemical cell size (top and center rows). 
Notably, in the case of the larger chemical cell size, a superposition of $\left<100\right>$ and $\left<110\right>$ orientations are observed.
This is consistent with experimental observations of ``nested'' sub-cell structures \cite{Bertsch2020, Depinoy2021,SMITH2019728}.
In contrast, the absence of cellular segregation leads to a diffuse autocorrelation function without obvious periodicity (bottom row).

Figure\,\ref{fig:Figure_4}c shows the average local dislocation density relative to the full volume average dislocation density along the $[001]$ crystal direction.
A global structural alignment of high dislocation density areas parallel to $\left<001\right>$ is evident only for DDD simulations with cellular segregation (top and center rows).
No obvious alignment in the $\left<100\right>$ direction is observed in the absence of cellular segregation (bottom row).
An averaging of the dislocation density along the $[101]$ direction, i.e. the intersection line between the $(\bar{1}11)$ and the $(\bar{1}\bar{1}1)$ slip planes, is shown in Fig.\,\ref{fig:Figure_4}d.
Thereby, dislocation structures on parallel slip planes can be identified, as shown for the case without cellular segregation in Fig.\,\ref{fig:Figure_4}d (bottom).
With cellular segregation, a periodic equidistant pattern corresponding to cell sizes of 800\,nm (top) and 1600\,nm (center) is observed parallel to the $(010)$ plane along with dislocation alignment parallel to $\{111\}$-planes extending between $\left<100\right>$-aligned dislocation walls. 
Notably, the out-of-plane spacing of the dislocation structures replicating $\{111\}$-planes is independent of the segregation cell size.

The current large-scale 3D DDD simulations demonstrate that dislocation structure orientations parallel to $\left<100\right>$-type directions can only form when cellular solute segregation is present. 
This effect is likely driven by Mo due to its large atomic misfit volume  \cite{antillonStructuralPointdefectCalculations2021,Senkov2001a}.
The regular periodic pattern of the $\left<100\right>$-type structures in accordance with the prescribed segregation cells closely resemble common experimental observations of dislocation cellular structures in connection with solute segregation shown in Fig.\,\ref{fig:Figure_1}c (bottom) and in the literature \cite{Wang2018,Liu2018,Bertsch2020,Voisin2021}.
This conclusion holds also for microstructure sections extracted normal to the $[100]$ crystal direction, shown in supplementary Fig.\,S5, where longitudinal dislocation alignment is evident for existing solidification cells.
In contrast, non-periodic dislocation alignment parallel to $\{111\}$-planes are in agreement with the dislocation structure observed experimentally in areas without solute segregation cells as in Fig.\,\ref{fig:Figure_1}c (top) and serve as an explanation for the origin of dislocation structures observed in additively manufactured pure metals \cite{Wang2020f}.

It should be noted here that estimations of the change in chemical composition due to solute diffusion induced by the internal stresses and thermal stress-gradients in the simulation without cellular solidification indicate no relevant influence of substitutional solute diffusion even at high temperatures as discussed in supplementary section S1.1.
Consequently, we conclude that previous suggestions based on solid diffusion, e.g. in \cite{Birnbaum2019}, as well as thermo-mechanical stresses alone cannot explain the well defined pattern-type dislocation structure alignment along $\left<100\right>$-axes.
However, it is also evident that different dislocation structure orientations can co-exist even when cellular solute segregation is present, which indicate a competition between dislocation multiplication and dislocation-solute interaction leading to the formation of both alignments.

\subsection{Coexistence of different dislocation structure alignments in a single grain}

\begin{figure}[htbp]
    \centering
    \includegraphics[width=\textwidth]{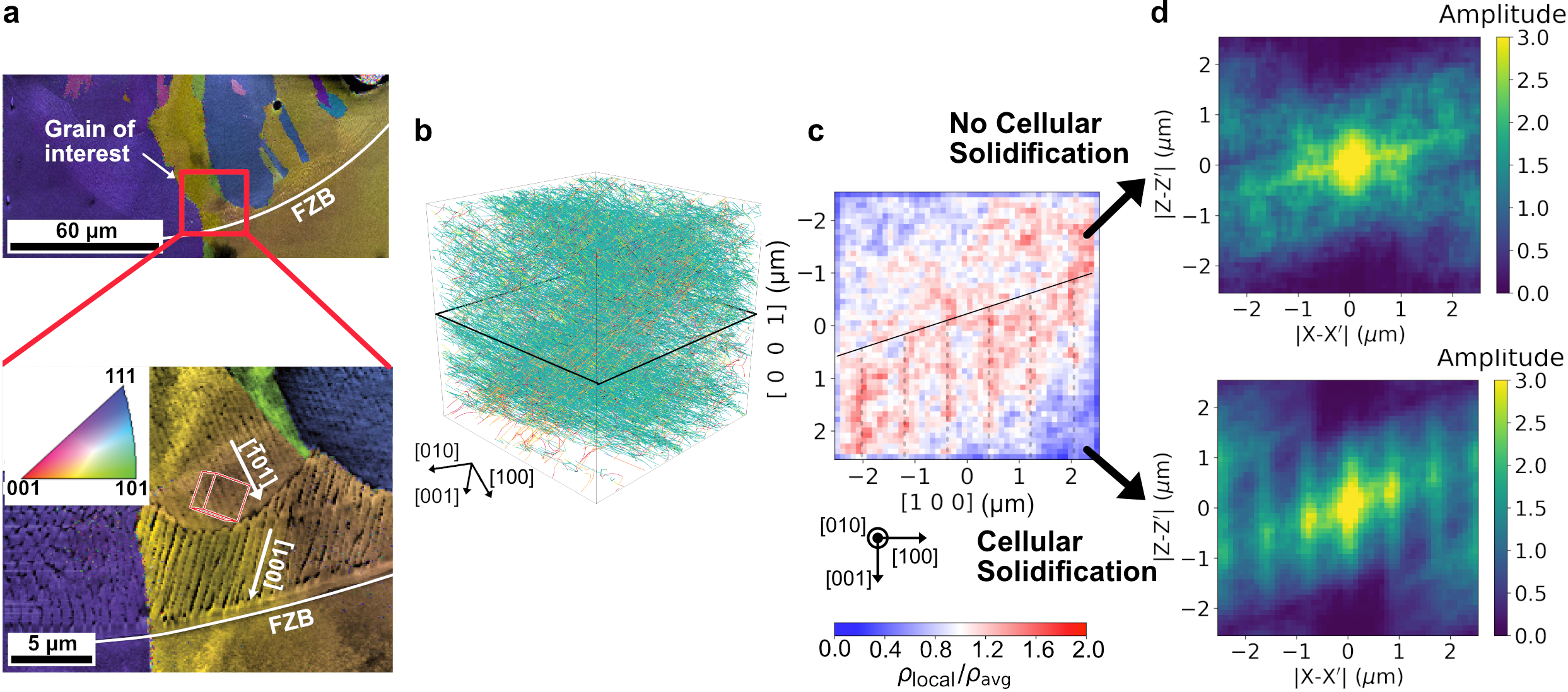}
    \caption{\textbf{Dislocation structure alignment in a single grain having both a cellular and non-cellular solidification structure.} 
    \textbf{a}, Low (top) and high (bottom) resolution EBSD-IPF maps of a grain with both $\left<001\right>$ and $\left<110\right>$-type surface structures (see also Fig.\,\ref{fig:Figure_1}).
    \textbf{b} Predicted dislocation structure at 700K after $\varepsilon_\mathrm{pl}^\mathrm{VM} = 0.85\%$ from the 3D DDD simulations, mimicking the crystal orientation of the grain shown shown in \textbf{a}.
    The plane indicated in black represents schematically the separation between the segregated and unsegregated domain.    
    \textbf{c}, Average local dislocation density relative to the full volume average density along the $[010]$-direction extracted from a section of the dislocation microstructure incorporating both solidification cells and unsegregated domain. Dashed grey lines indicate the position of the center plane of the solute segregation walls along the $[100]$-direction and the black line  represents the intersection of the plane separating the segregated and segregated domains.
    \textbf{d}, 2D contour-plots showing the amplitude of autocorrelation functions using the dislocation density maps in \textbf{c} isolating the unsegregated (bottom) from the segregated (top) domain.}
    \label{fig:Figure_5}
\end{figure}

Figure\,\ref{fig:Figure_5}a shows a high magnification, EBSD IPF map for a grain close to the FZB (see also Fig.\,\ref{fig:Figure_1}) containing cellular surface structures aligned with a $[001]$ direction solute segregation-induced structure, adjacent to a region exhibiting $[\bar{1}01]$ aligned features (dislocation interaction-induced structure). Key measures such as the crystal orientation and the alignment of the solidification cells as well as the residual stress as computed from the FEM simulations (Fig.\,\ref{fig:Figure_2}) are used as inputs for a 3D DDD simulation at 700K. In the DDD simulation, a cubic simulation cell is chosen with edge length of $10\,\upmu$m and a heterogeneous chemical distribution, i.e. cellular segregation, is assumed in the lower half of the simulation volume, while the upper half is assumed to have a uniform composition, see Fig.\,\ref{fig:Geometry}d. 

The evolution of the dislocation ensembles driven by the residual stresses corresponding to the cool-down stage at 700K leads to the formation of a dense dislocation microstructure in the center of the simulation domain with an average dislocation density of $\rho \approx 4.9 \times 10^{13}\,\mathrm{m}^{-2}$, shown in Fig.\,\ref{fig:Figure_5}b.
Here, the separating plane between unsegregated (upper half) and segregated (lower half) domain is visualized by black lines.
The average local dislocation density relative to the full volume average dislocation density along the $[010]$ crystal direction incorporating both the segregated and unsegregated domain is shown in Fig.\,\ref{fig:Figure_5}c, where the black line visualizes the boundary between the segregated and unsegregated domain.
A higher dislocation density is observed to align with the position of the segregation cell walls indicated by the dashed grey lines in the bottom half of the simulation volume, whereas no clear dislocation structure alignment is evident in the unsegregated domain in the upper half of the volume.
Clearly, the emergence of periodic alignment of high dislocation density parallel to the $[001]$-direction is evident only in the domain containing cellular solidification.
Corresponding 2D autocorrelation functions shown in Fig.\,\ref{fig:Figure_5}d for the segregated (bottom) and unsegregated (top) domains using locations of high dislocation density provide for a closer study of the difference in the morphology of the patterns.
Whereas for the segregated section distinct peaks in the autocorrelation function are evident with a distance corresponding to the 800\,nm solidification cell wall spacing along the $X$-direction (corresponding to $[100]$), a rather diffuse pattern with alignments closer to typical fcc-directions are observed in the case of the unsegregated section.

These DDD simulations mimicking an experimental grain orientation confirm the role of cellular segregation on the formation of two different microstructural morphologies within the same grain.
The similarity of the observed dislocation alignments in $\left<100\right>$ directions with previous results discussed in Fig.\,\ref{fig:Figure_3} and \ref{fig:Figure_4}, which incorporate a grain orientation in perfect alignment with laser scanning and build directions, indicate that the formation of dislocation subcells is independent of individual grain orientations.
Furthermore, it is evident that solute segregation-driven dislocation walls do not progress into areas without solute segregation.
This shows that the dislocation cell walls commonly observed experimentally in LPBF processed AISI 316L SS form as a product of solute-dislocation interaction.

The current simulations and experimental of single-tack scans do not account for cyclic heating, which would occur in bulk AM processing. However, the fact that a clear cellular dislocation microstructure is observed in the present single-track scans indicates that while bulk processing effects may \textit{contribute} to the formation of this characteristic microstructure, they are likely not a factor in explaining its formation on a fundamental level.
In this respect, it has to be noted that the current DDD simulations only rely on the influence of the atomic misfit volume on the dislocation CRSS. More accurate chemical composition-dependent parameters for cross-slip, dislocation drag coefficient, and possibly elastic constants and lattice constants will allow for a more accurate representation of the influence of microsegregation on dislocation mobility. However, by accounting for more composition-dependent effects, the observed mechanisms are likely to be even more amplified.

\section{Conclusion}

Experimental characterizations of the underlying dislocation microstructure of single track LPBF scans in AISI 316L stainless steel are coupled with macro-scale finite element analyses of the evolving transient thermo-mechanical stresses and large scale three-dimensional discrete dislocation dynamics simulations of the dislocation structure evolution during the cool-down phase after the single track scan. 
The results of this work show the relationship between the pre-existence of cellular segregation of heavier elements such as Mo and Cr in solidification cell walls during LPBF processing of AISI 316L SS and the formation of dislocation wall patterns aligned parallel to the $\left<100\right>$ crystal orientations.
Although the significant residual stresses emerging from restriction of contraction during rapid cool-down lead to the formation of diffuse dislocation cells in agreement with experimental observations in absence of solidification cells, those dislocation structures significantly differ from the pattern-like $\left<100\right>$-aligned dislocation cells, which form in conjunction with solidification cells even in single LPBF tracks.
It can therefore be concluded that mechanisms based on thermo-mechanical stresses alone cannot provide a sufficient explanation for the formation of well-defined cellular structures observed in LPBF of AISI 316L.
Rather, it is evident from large scale 3D DDD simulations that observations of $\left<100\right>$-aligned dislocation walls are driven by an interruption of dislocation slip by segregated solute atoms, leading to enhanced cross-slip and dislocation interaction in later stages of the cool-down process.
This conclusion is in agreement with experimental observations of cellular dislocation structures in domains incorporating chemical segregation in the current single track scans.

Furthermore, we show that $\left<100\right>$-type and $\left<110\right>$-type dislocation structure alignments can occur simultaneously in the same grains.
This is evident from experimental and DDD observations of domains incorporating both cellular and non-cellular solidification structures.
The existence of those two separated domains within the same grain may be a feature of single-track LPBF scans in which bulk-processing effects related to cyclic heating are not present.
However, the DDD observation of ``nested'' sub-cell structures in case of larger solidification cells is in agreement with bulk-processing experiments \cite{Bertsch2020, Depinoy2021,SMITH2019728}.
This coexistence of different alignments originates from a competition between plastic slip on glide planes and pinning effects of dislocations at sites of enhanced slip resistance induced by segregation of larger elements leading to a formation of well-aligned dislocation walls.
The interplay of dislocation slip and pinning effects suggests that unique mechanical properties observed experimentally, such as anisotropy \cite{Chen2019}, temperature-dependency \cite{Zhong2016}, or a combination of high strength and ductility \cite{Liu2018,Wang2018} are largely determined by an interplay between dislocation slip and cellular segregation.
Although the current study is limited to AISI 316L SS, it is therefore likely that the conclusion can at least partly be transferred to other alloys showing microsegregation of larger elements such as Nickel-base superalloys \cite{zhang2017homogenization}.

This work also provides the basis for a reliable prediction of mechanical properties of LPBF processed alloys by identifying a mechanistic perspective on the origin of the dislocation microstructure induced by residual stresses and cellular solidification.
The understanding of the interactions between residual stresses and solute segregation will enable the development of unique physically informed crystal plasticity simulations.
For example, the predicted cellular dislocation microstructure could be used to estimate the heterogeneous internal stress profiles, and thus, provide a more physical basis for the incorporation of strengthening effects of cellular structures into continuum scale models. Further, the dislocation microstructure evolution itself can be used to inform dislocation-based continuum theories of plasticity. Thereby, this work can enable designing new additively manufactured materials with superior mechanical properties through a multi-scale materials modeling framework.

\section*{Acknowledgments}
 
The authors gratefully acknowledge support for this work from the Office of Naval Research through the Naval Research Laboratory’s core funding and grant number N00014-18-1-2858 as well as by the U.S. National Science Foundation (NSF) CAREER award CMMI-1454072.
Some simulations were conducted at the Advanced Research Computing at Hopkins (ARCH) core facility (rockfish.jhu.edu), which is supported by an NSF grant number OAC-1920103. Some simulations were also conducted using the Extreme Science and Engineering Discovery Environment (XSEDE) Expanse supercomputer at the San Diego Supercomputer Center (SDSC) through allocation TG-MAT210003. XSEDE is supported by National Science Foundation grant number ACI-1548562.

\section*{Competing interests}
\noindent 
The authors declare no competing interests.

\section*{Author contributions}
\noindent 
\textbf{Jaafar El-Awady} designed and led the computational aspects of this project.
\textbf{John Michopoulos and Andrew Birnbaum} designed and led the experimental aspects of the project.
\textbf{Andrew Birnbaum} performed the experiments.
\textbf{Patrick Callahan} performed the TEM imaging. 
\textbf{Markus Sudmanns, Athanasios Iliopoulos and John Michopoulos} developed, conducted and evaluated the FEM simulations.
\textbf{Markus Sudmanns, Yejun Gu and Jaafar El-Awady} developed and evaluated the 3D DDD simulations.
All authors contributed to the analysis and discussion of the results.
\textbf{Markus Sudmanns, Andrew Birnbaum and Jaafar El-Awady} wrote the initial manuscript and all co-authors provided input to the final manuscript.

\section*{Supplementary materials}
\noindent 
Supplementary Methods and Materials\\ 
Supplementary Figures and Tables\\
Supplementary Movies




\bibliographystyle{unsrt}

\end{document}





\title{Supplementary materials for: \\ The interplay of local chemistry and plasticity in controlling microstructure formation during laser powder bed fusion of metals}

\author{Markus Sudmanns$^{1}$, Andrew J. Birnbaum$^{2}$, Yejun Gu$^{3,1}$, \\ Athanasios P. Iliopoulos$^{2}$, Patrick G. Callahan$^{2}$,\\ John G. Michopoulos$^{2}$, Jaafar A. El-Awady$^{1\ast}$\\
\\
\normalsize{$^{1}$Department of Mechanical Engineering, The Whiting School of Engineering,}\\\normalsize{The Johns Hopkins University, Baltimore, MD 21218, USA}\\
\normalsize{$^{2}$United States Naval Research Laboratory, Washington DC, 20375, USA}\\
\normalsize{$^{3}$Institute of High Performance Computing, A*STAR,} \\\normalsize{Fusionopolis, 138632, Singapore}\\
}

\date{}



\maketitle

\renewcommand{\thesection}{S \arabic{section}}
\renewcommand{\thesubsection}{S 1.\arabic{subsection}}

\section{Materials and Methods}

\subsection{Autocorrelation analyses of dislocation density walls in DDD simulations}

Two-dimensional (2D) spatial correlation functions were used to provide an identification and a quantitative analysis of periodic patterns in the dislocation microstructure.
To prepare the data for the $5^3\,\upmu$m$^3$ simulations at 1300K, 1100K, 900K and 700K (Fig.\,4), local dislocation densities were calculated by discretizing the full simulation volume, i.e. $5\times5\times5\upmu$m$^3$, into $50\times50\times50$ voxels. 
The respective average dislocation density was computed by averaging along the $[001]$ and $[100]$ direction for each simulation, respectively, to create a 2D image.
For the $10^3\,\upmu$m$^3$ 3D DDD simulations with 800\,nm and 1600\,nm solidification cell wall spacing and the $5^3\,\upmu$m$^3$ simulation in the absence of cellular segregation at 700K (Fig.\,5), local dislocation densities were calculated by extracting a 400\,nm thin section of the 3D volume discretized into $100^3\,$nm$^3$ sized cubic voxels. 
Thereby, sub-volumes of $10\times10\times0.4\,\upmu$m$^3$ and $5\times5\times0.4\,\upmu$m$^3$ size were generated, consisting of $100\times100\times4$ voxels and $50\times50\times4$ voxels, respectively. 
Then, the density was averaged along the 400\,nm thick section in $[001]$ crystal direction (consisting of 4 voxels).
For the simulation mimicking an experimental grain orientation (Fig.\,6), volume averages of the dislocation density along the $[010]$ direction were extracted from a sub-volume of size $5^3\,\upmu$m$^3$ in the center of the simulation volume to limit the influence of boundaries on the evaluation.
Since only features of high dislocation density are of interest for the study of dislocation patterns, we only considered local dislocation densities in which contain a density greater than the $1.15 \times$ the full volume average for the calculation of the autocorrelation function in the simulations at different temperatures (Fig.\,4) and the simulation mimicking experimental grain orientation (Fig.\,6).
Further, each local dislocation density value was divided by the full volume average to obtain a relative measure of dislocation density accumulation.
This method creates a data-set for each simulation consisting of $100\times100$ (in case of the $10^3\,\upmu$m$^3$ simulations) and $50\times50$ (in case of the $5^3\,\upmu$m$^3$ simulations and in case of the simulation with experimental grain orientation) data points containing the local relative dislocation density.

Using these data-sets, we calculated the spatial autocorrelation functions, which are often used in digital image processing to correlate one image with itself \cite{jahne2005digital, HEILBRONNER1992351}. 
The autocorrelation function describes the similarity between a data-set and a copy of itself, which is shifted in two directions described by a vector $\mathbf{u}$ as a function of the distance $u = |\mathbf{u}|$.
For an efficient calculation of the autocorrelation function, we make use of the relationship between the autocorrelation function and the Fourier transform, known as the Wiener-Khinchin theorem \cite{wiener1930generalized, khintchine1934korrelationstheorie}.
This theorem states that the autocorrelation function of a stationary signal is equal to the Fourier transform of the power spectral density.
The autocorrelation function $C(\mathbf{u})$ of a signal $X$ (i.e. the generated data-set in case of our analysis) is then given by
\begin{equation}
C(\mathbf{u}) = E^{-1}\left[E(X)\bar{E}(X)\right], 
\end{equation}
where $E(X)$ is the Fourier transform and $\bar{E}(X)$ the complex conjugate.

\subsection{Estimation of solute diffusion}
Here we study the solute diffusion in $Cr_{17.5}Mo_{2.3}Fe_{80.2}$. This ternary material is sufficiently representative to well approximate the diffusion process in AISI 316L stainless steels since they have very similar chemical compositions (see Table~1). Taking all constituting elements in AISI 316L SS into account would require a significantly greater effort without changing the overall conclusion.

In SLMed AISI 316L stainless steels, dislocations can act as perfect vacancy/interstitial sinks and sources since their density is high. With the perfect sink assumption, the vacancy site fraction, $\chi_v$, is always in equilibrium in response to the hydrostatic stress $\sigma_H$:
\begin{align}
\chi_v=\chi_{v0}\exp{\left(\frac{\bar{\Omega}\sigma_H}{RT}\right)},
\end{align}
where $\chi_{v0}$ is the equilibrium vacancy site fraction, $R$ is the gas constant, and $T$ is the temperature.

According to the vacancy mediated substitutional diffusion theory~\cite{van2010,fischer2014}, the diffusion equation for the element $k$ is:
\begin{equation}
\begin{split}
\frac{\partial \chi_k}{\partial t}=
&D_k\nabla\cdot\left[\nabla\chi_k+\frac{1-f}{f}\left(\sum_{j=1}^n\Tilde{D}_j\nabla \chi_j\right)\right]\\
&-\chi_k\nabla\cdot\left[\frac{1}{f}\sum_{j=1}^nD_j\left(\nabla\chi_j-\chi_j\frac{\Omega_j\nabla\sigma_H}{RT}\right)\right]\\
&-\chi_k\exp{\left(\frac{\bar{\Omega}\sigma_H}{RT}\right)}\frac{\chi_v\bar{\Omega}\cdot\partial\sigma_H/\partial t}{RT},
\end{split}
\end{equation}
where $D_k$ is the diffusion coefficient of the element $k$, $\Tilde{D}_j=\frac{D_j}{\sum_{i=1}^{n}\chi_iD_i}$ is the relative diffusion coefficient, $f$ is the correlation factor, $\Omega_j$ is the molar volume of the element $k$ and the weighted average molar volume is $\bar{\Omega}=\sum_{i=1}^n \chi_i\Omega_i$. 

The parameters in Table~\ref{tab:diffusion_param} and the dislocation microstructure at $t=2.437\times10^{-3}s$ are used in the calculations of substitutional diffusion in SLMed AISI 316L stainless steels. The results show that even at high temperature and with high stress field, the substitutional diffusion can hardly occur. Therefore, the chemical segregation observed in AM experiments~\cite{Wang2018,Birnbaum2019} is likely dictated by the constitutional supercooling. 

\section{Supplementary Figures and Tables}

\renewcommand{\thetable}{S\arabic{table}}
\setcounter{table}{0}

\renewcommand{\thefigure}{S\arabic{figure}}
\setcounter{figure}{0}

\begin{figure}[H]
    \centering
    \includegraphics[width=0.5\textwidth]{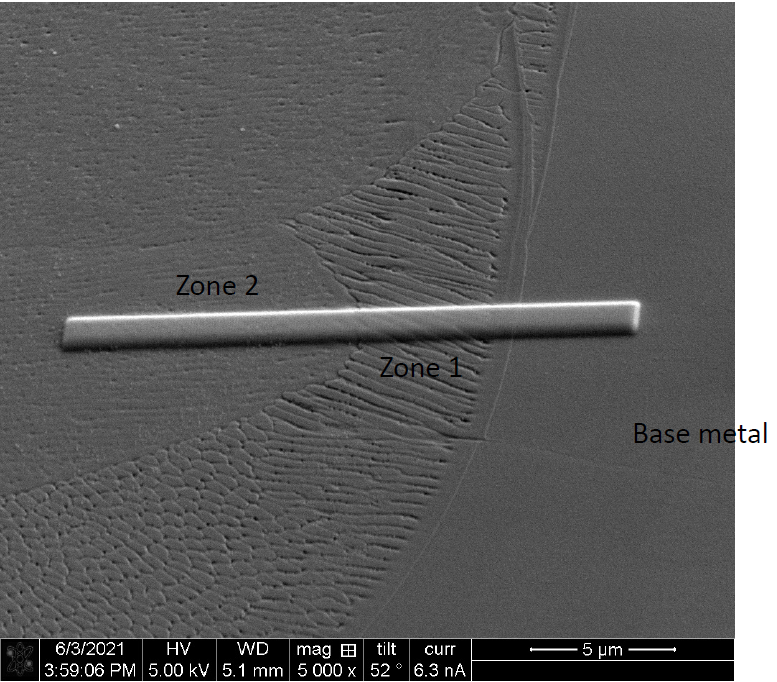}
    \caption{SEM image detailing the location of the sample prepared with focused ion beam (FIB) through the unsegregated area (Zone 2), the segregated area (Zone 1) and the base metal for extracting the TEM images shown in Fig.\,1c.}
    \label{fig:TEM_foil_location}
\end{figure}

\begin{figure}[H]
    \centering
    \includegraphics[width=\textwidth]{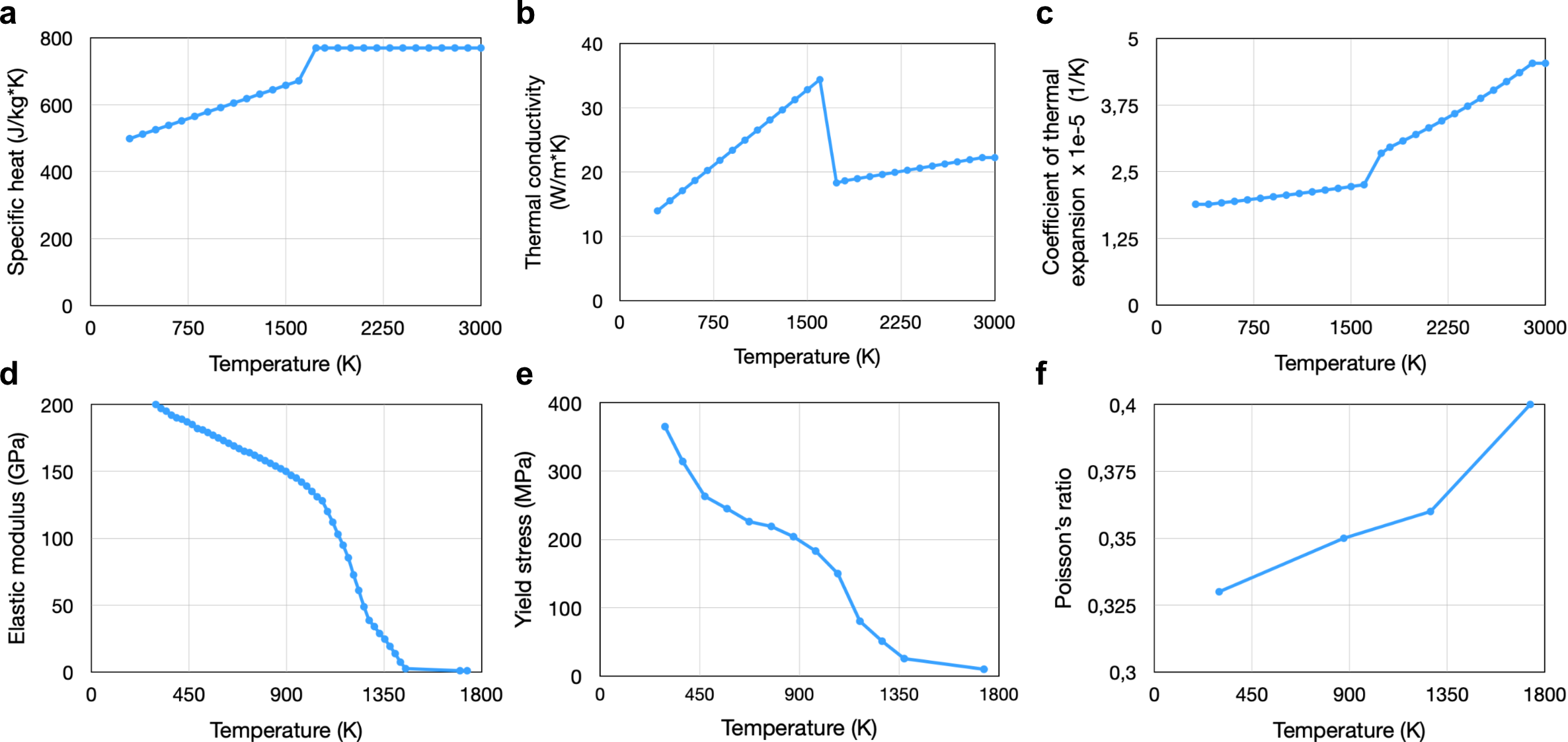}
    \caption{Temperature-dependent material parameters compiled from literature \cite{ChaudhuryAtlas, desu2016mechanical, byun25854mechanical} as used in the FEM and DDD simulations. \textbf{a}, Specific heat, \textbf{b}, Thermal conductivity, \textbf{c}, Coefficient of thermal expansion, \textbf{d}, Elastic modulus, \textbf{e}, Yield stress, \textbf{f}, Poisson's ratio. }
    \label{fig:material_parameters}
\end{figure}
\begin{figure}[H]
    \centering
    \includegraphics[width=\textwidth]{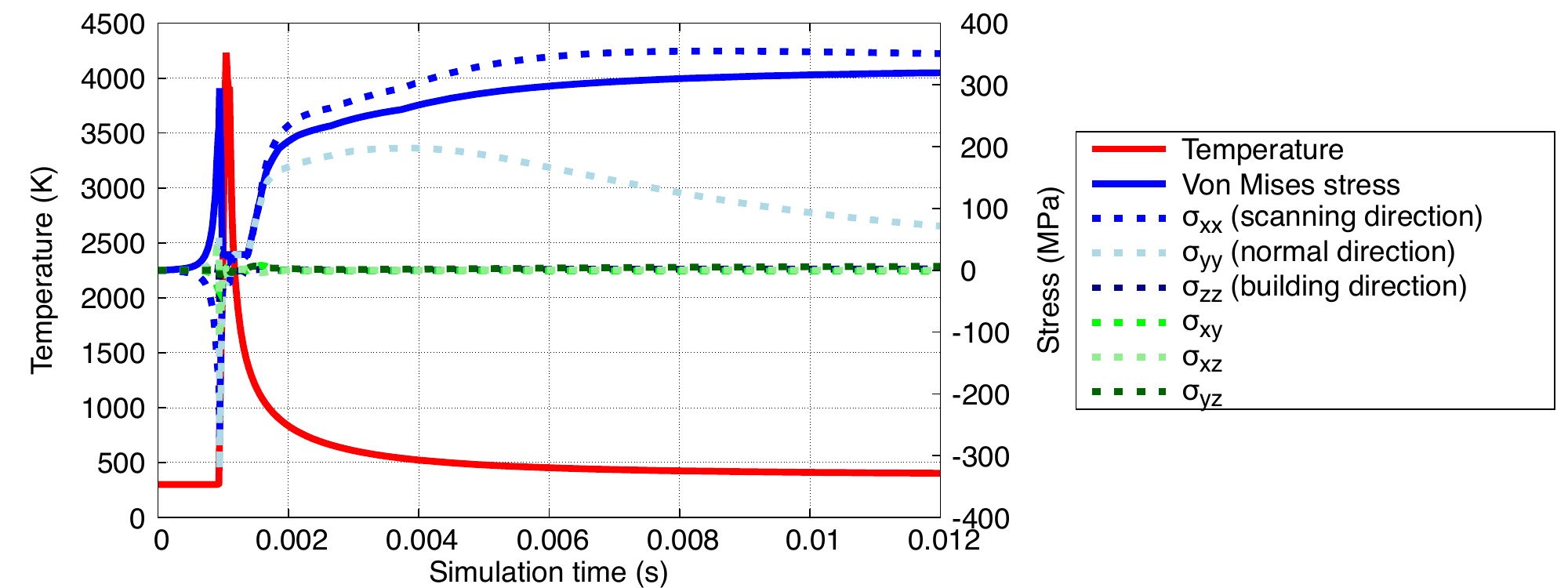}
    \caption{Evolution of the Temperature, Von Mises stress, and stress components in the FEM simulation of SLM cooldown.}
    \label{fig:Stress_components}
\end{figure}
\begin{figure}[H]
    \centering
    \includegraphics[width=\textwidth]{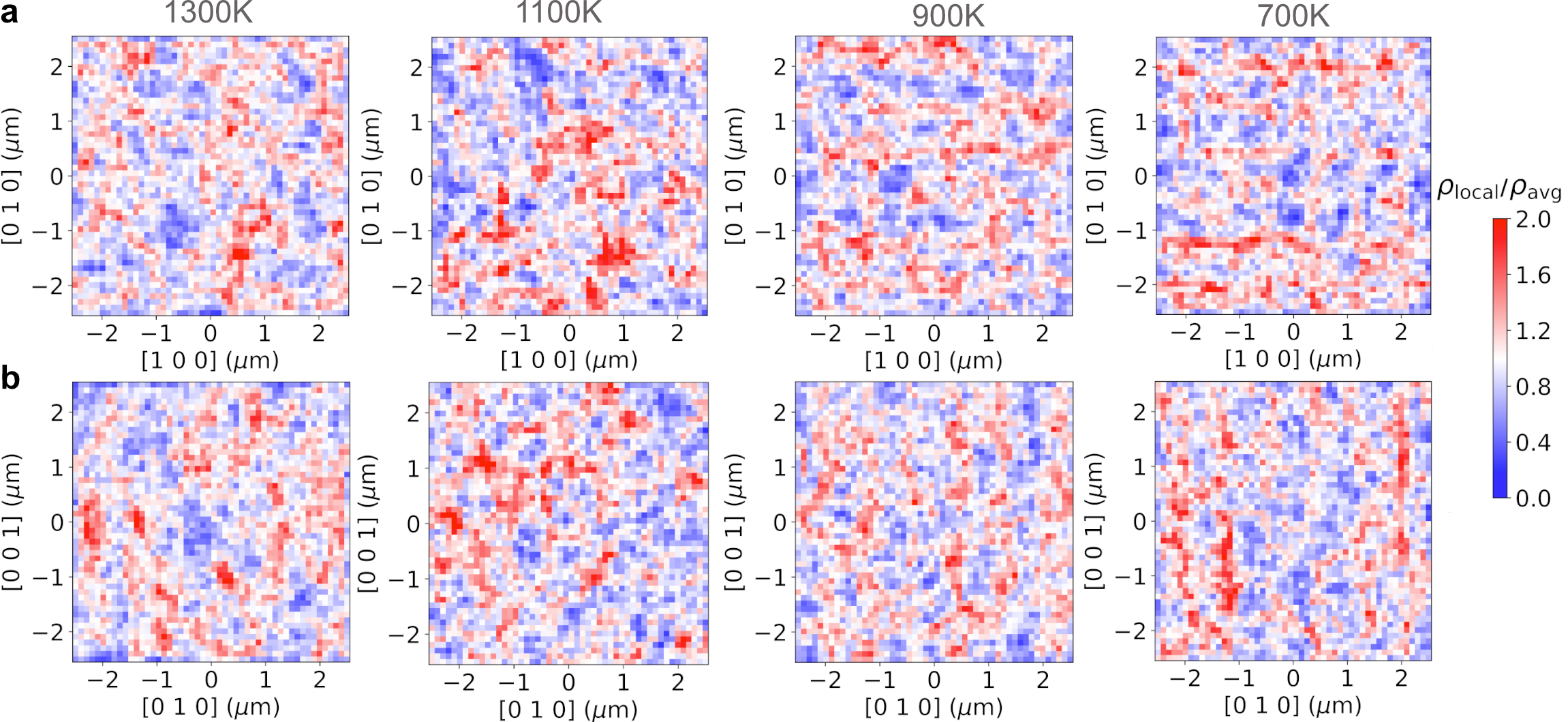}
    \caption{The average local dislocation density along the: \textbf{a}, $[001]$, and \textbf{b}, $[100]$ crystal directions relative to the full volume average density from four $5^3\,\upmu$m$^3$ 3D DDD simulations at 1300K, 1100K, 900K and 700K with cellular segregation with a cell wall spacing of 800\,nm according to the experimental observations. See Fig.\,1c for the geometry of the solidification cell walls.
    }
    \label{fig:Temperature_dependency}
\end{figure}
\begin{figure}[H]
    \centering
    \includegraphics[width=\textwidth]{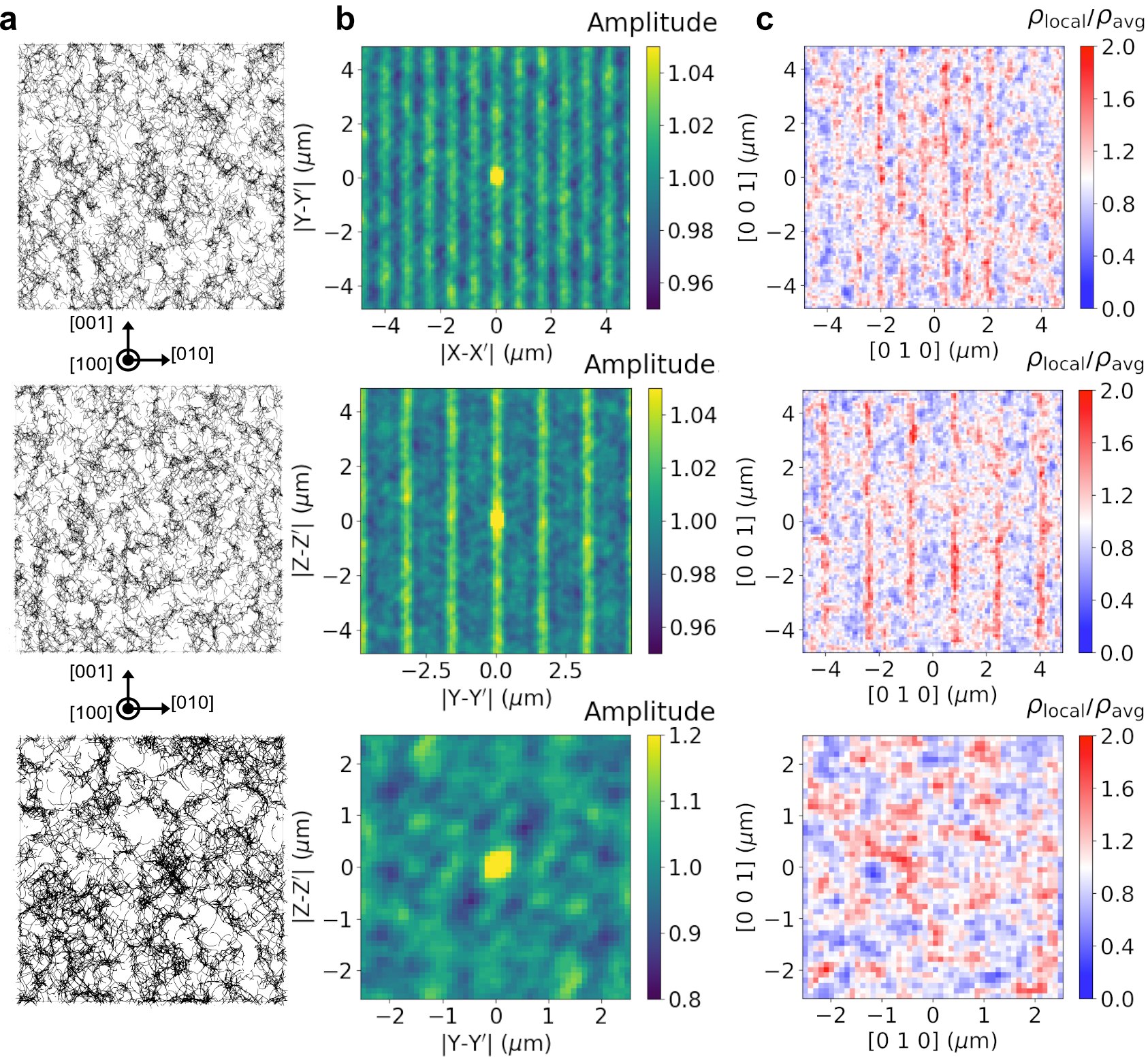}
    \caption{$10^3\,\upmu$m$^3$ 3D DDD simulations assuming 800nm (top row) and 1600nm solidification cell size (center row) and a $5^3\,\upmu$m$^3$ 3D DDD simulation in the absence of cellular segregation (bottom row). All simulations are at 700K and subjected to the stress state obtained from the FEM simulations. 
    \textbf{a}, The predicted dislocation microstructures in 400nm thick sections extracted from DDD normal to the $[100]$ direction (see coordinate system for reference). 
    \textbf{b}, Autocorrelation analyses using an average local dislocation density map extracted from sections shown in \textbf{a}, where the $Y$- and $Z$-axes correspond to the $[010]$ and $[001]$ crystal directions, respectively.     
    \textbf{c}, Average local dislocation density $\rho_\mathrm{local}$ relative to the full volume average density $\rho_\mathrm{avg}$ along the $[100]$-direction. 
    }
    \label{fig:Orientations}
\end{figure}

\begin{table}[H]
\centering
\caption{Material parameters for the substitutional diffusion calculations}
\begin{tabular}{cc}
\hline
Quantity & Value \\\hline
$f$& 0.781~\cite{burton1975diffusion}\\
$V_{Cr}$ &  $7.23\times10^{-6} m^3/mol$\\
$V_{Mo}$ &  $9.4\times10^{-6} m^3/mol$\\
$V_{Fe}$ &  $7.1\times10^{-6} m^3/mol$\\
$D_{Cr}$ &   $6.3\times10^{-6}\exp{\left(-\frac{243 kJ/mol}{RT}\right)}$~\cite{Smith1975}\\
$D_{Mo}$ &   $3.6\times10^{-6}\exp{\left(-\frac{239.8 kJ/mol}{RT}\right)}$~\cite{Alberry1974c}\\
$D_{Fe}$ &   $1.18\times10^{-6}\exp{\left(-\frac{228.5 kJ/mol}{RT}\right)}$~\cite{Patil1982}\\
$\partial \sigma_H/\partial t$ & $1.0\times 10^{10} MPa/s$\\
$\chi_{v0}$ & 0.1\%\\\hline
\label{tab:diffusion_param}
\end{tabular}
\end{table}

\renewcommand\refname{Supplemental References}





\bibliographystyle{unsrt}